\documentclass[%
 reprint,
 amsmath,amssymb,
 aps,
]{revtex4-2}

\usepackage{graphicx}
\usepackage{dcolumn}
\usepackage{bm}

\usepackage[utf8]{inputenc}
\usepackage{amssymb, amsmath}
\makeatletter
\def\amsbb{\use@mathgroup \M@U \symAMSb}
\makeatother
\usepackage[dvipsnames]{xcolor}
\def\bpsi{{\bf \psi}}
\def\hrho{\hat{\rho}}
\def\rv{{\bf r}}

\def\br{{\bf r}}

\def\bv{{\bf v}}

\def\bk{{\bf k}}

\def\bv{{\bf v}}
\def\nn{\nonumber}
\def\nn{\nonumber}

\newcommand{\commentout}[1]{}

\begin{document}

\preprint{APS/123-QED}

\title{Theory of nonlinear optical response}

\author{Amin Maleki Sheikhabadi}
\affiliation{%
 Department of Physics, Shahid Beheshti University, Evin, Tehran 1983969411, Iran 
}%


\author{Zahra Bagheri}
\affiliation{
School of Particles and Accelerators, Institute for Research in Fundamental Sciences (IPM), P.O. Box 19395-5531, Tehran, Iran
}%

\author{Ali Sadeghi}
\affiliation{%
  Department of Physics, Shahid Beheshti University, Evin, Tehran 1983969411, Iran
}%



\begin{abstract}

We present a general formalism for investigating the second-order optical response of solids to an electric field in weakly disordered crystals with arbitrarily complicated band structures based on density-matrix equations of motion, on a Born approximation treatment of disorder, and on an expansion in scattering rate to leading non-trivial order. One of the principal aims of our work is to enable extensive transport theory applications that accounts fully for the interplay between electric-field-induced interband and intraband coherence, and Bloch-state scattering. The quasiparticle bands are treated in a completely general manner that allows for arbitrary forms of the intrinsic spin-orbit coupling (SOC) and could be extended to the extrinsic SOC. According to the previous results, in the presence of the disorder potential, the interband response in conductors in addition to an intrinsic contribution due to the entire Fermi sea that captures, among other effects, the Berry curvature contribution to wave-packet dynamics includes an anomalous contribution caused by scattering that is sensitive to the presence of the Fermi surface. To demonstrate the rich physics captured by our theory, the relaxation time matrix for different strength order is considered and at the same time  we explicitly solve for some electric-field response properties of simple disordered Rashba model that are known to be dominated by interband coherence contributions. The expressions we present are amenable for numerical calculations, and we demonstrate this by performing a full band-structure calculation of the interband contribution, even in metals.

\end{abstract}

\maketitle


\section{Introduction}

The study of steady state response of a solids to external electric fields is subject of intensive theoretical and experimental investigations Refs.~\cite{boyd1992,kai2016,fridkin2001}, but there are no reasons why this effect would not consider in a more complex form like in metals~\cite{Osterhoudt2019,Neupane2016,McIver2012,Andrew2017}. Originally the effect was first observed in the very intensive laser light in 60s allowed irradiated systems to be taken out of equilibrium that currently investigates to many electrical applications~\cite{boyd2003}. The optical response of solids is fundamental theory of the electric polarizations based on polarization suseptibilities~\cite{Rabe2007book}, characterizing not only linear response but also nonlinear of the materials to the electromagnetic field~\cite{boyd1992}. Roughly speaking, although the susceptibilities determine all the optical response of solids, but recently we have tarted to probe this information generate by all the optical response of metals. An example of these applications is to design of novel photovoltaic cells~\cite{power2016,Yevgeny2017,Andrew2017}. The reason for not examining this effect in metallic systems is that the steady-state response of a metal or a doped semiconductor to external magnetic field is complicated in the band structure ~\cite{culcer2017interband,Culcer2017} and also the essential roles are played by host crystal due to disorder or thermal excitation of lattice degree-of-freedom~\cite{vasko2006quantum}. Currently, fascinating ongoing experiments start to observe this effect in the surface state of topological insulators and Weyl semimetals~\cite{Yevgeny2017,Osterhoudt2019,yang2017,McIver2012}. 

In this present work, we start with supposing that the Bloch state scattering mechanisms that plays dominate role in the limit of the repopulation of states close to the Fermi level in the direction of the electric field. The require consideration to use Boltzmann equation is to assume that the dominate Bloch state scattering mechanisms is weak~\cite{allen1978} because of the finite size of the system including many interfaces with strong spin-orbit splitting, including metals with semiconductor giant SOC or insulators~\cite{Karube2016,Chen2016,aminsu2}. To do so, by using quantum kinetic equation, we present a general formulation to describe the Bloch state density matrix. The diagonal and off-diagonal terms in density matrix are responsible for charge and spin density to spatially uniform and constant field which appears at the leading sub-dominate order in a weak scattering expansion. To the second order response, the diagonal term in density matrix which is responsible for longitudinal response dominates and captured by the relaxation time matrix. However the off-diagonal terms in density matrix is arising from interband coherence terms that contains two major parts, band structure and off-diagonal terms~\cite{culcer2017interband,Culcer2017}, coming from the entire Fermi surface including the Berry phase contributions and scattering anomalous deriving term on the Fermi surface. Hence we analytically derive both parts by using the Boltzmann equation and practically show that which part under what conditions can be dominated. Generally we present a general quantum kinetic theory of linear and nonlinear response to an electric field applied to any model systems that requires an approach working for any band structure with arbitrarily complicated band structures including the interband and interaband coherence response and the Bloch-state repopulation responses on an equal footing. Although this theory is a very powerful approach, it is very useful to also show how the same results can be obtained independently by using the different methods such as the SU(2) gauge theory formulation~\cite{aminsu2}, diagrammatic approach of the Kubo linear response theory~\cite{kubo,amincubic}.

The outline of the rest of the paper follows. In the section [\ref{dipolehamiltonian}], we consider the simplest approach, known as ``independent particle approximation'' to neglect the electron-electron interaction except insofar as it is included in the calculation of the band structure of the solid. Clearly, quasiparticle corrections and exciton effects associated with the creation of electron-hole pairs, as well as local field corrections, should be considered in this approach.  In section [\ref{currentdensity1}], we develop a perturbation expansion in the case of the Born approximation treatment of disorder, and on an expansion in scattering rate to leading nontrivial order. The Bloch state density matrix of a crystal to a spatially constant electric field assumes that the Wannier representation Bloch Hamiltonian is present as a function of wave vector and treats as short range impurity potential. Analysis shows that the verity of different times in the approximations for disorder in crystal leads to the density matrix response. Then our calculation focuses to the case when only off-diagonal response of the Bloch state density matrix in metallic systems present and in the simple case of disordered Rashba model our theory is investigated~[\ref{disorderedRashba}]. Finally, we state our conclusions in Section \ref{conclusion}.

\section{THE DIPOLE HAMILTONIAN}\label{dipolehamiltonian}

Full many-particle Hamiltonian in the presence of disorder potential is
\begin{eqnarray}
	\hat{H}=\int {\boldmath\bpsi}^\dagger({\bf r}, t) \left( \frac{{k}^2}{2m}+ U_b(\rv )+V(\rv)\right)\bpsi({\bf r}, t)\label{hamiltonian},
\end{eqnarray} 
describing the Bloch electrons in disordered system. In Eq.~\ref{hamiltonian}, $U_b(\rv)$ is the periodic potential $U_b(\rv +a)$ with $a$ being any lattice vector, and ${k}=-i \hbar \nabla$ the momentum operator. In the simplest case of Gaussian white noise disorder potential, $V(\rv)$ is assumed to have zero average and Gaussian distribution given by $\langle V(\rv)V(\rv')\rangle=(\hbar/(2\pi N_0\tau_0))\delta(\rv-\rv')\rangle$, where $\langle \cdots \rangle$ denotes the angular average over the momentum direction, $N_0=m/2\hbar^2\pi$, $n_i$ and $v_0$ are the single-particle density of states per spin in the absence of SOC, the impurity concentration and the scattering amplitude, respectively. $\tau_0$ is the elastic scattering time at the level of the Fermi Golden Rule. From now on, we work with units such that $\hbar=c=1$, $e=-|e|$ for simplicity’s sake.

The homogeneous electric field is connected to the Hamiltonian with the minimal substitution, ${\bk } \rightarrow e {\bf A}(t)$, where ${\bf A}(t)$ is a vector potential that describes the (macroscopic Maxwell) electric field as ${\bf E}=\dot{\bf A}(t)$. Hence we define a new field operator in the  Heisenberg picture as
\begin{eqnarray}
	\psi(\rv,t) \rightarrow \tilde{\psi}(\rv,t) \equiv \psi(\rv,t)e^{-ie{\bf A}(t)\cdot\rv}
\end{eqnarray}
and it is easy to verify that in the Heisenberg picture it formally satisfies  the equations of motion
\begin{eqnarray}
	i\frac{d \tilde{\psi}(\rv,t) }{dt}=[\tilde{\psi}(\rv,t), H_{eff}(t)]
\end{eqnarray}
then after the gauge transformation, the dipole Hamiltonian is written as 
\begin{eqnarray}
	{H}_{eff}(t)=  {H}_0+{H}_{\bf E}(t)+{H}_{imp}({\bf r}),\label{hamiltonian}
\end{eqnarray}
where 
\begin{eqnarray}
	H_0=\int \tilde{\psi}^\dagger({\bf r}, t)\big(  \frac{{k}^2}{2m}+ U_b(\rv ) \big)\tilde{\psi}({\bf r}, t)
\end{eqnarray}
and the electric dependent part is given by 
\begin{eqnarray}\label{hamiltonian_p}
	{H}_{\bf E}=-e\int d\rv \bpsi^\dagger \rv \cdot {\bf E} \bpsi\equiv-{\bf P}\cdot {\bf E},
\end{eqnarray}
with ${\bf P}$ being the polarization operator.  From Bloch's theorem, the eigenfunction of ${H}_0$ can be chosen of the form 
\begin{eqnarray}
	\psi_n(\rv,t)=u_n({\bf k},\rv) e^{i{\bf k}\cdot\rv}
\end{eqnarray}
with the band index $n$ and crystal momentum $\bf k$, where ${ u}_n({\bf k},\rv)={ u}_n({\bf k},\rv+{\bf a})$ is the period part of the Bloch function. Hence in the Bloch states, the new field operation can be expanded as
\begin{eqnarray}
	\tilde{\psi}(\rv,t)=\sum_{n{\bk}}\bpsi_n(\rv,t) a_n(\bk)
\end{eqnarray}
with $ a_n^\dagger(\bf k)$ and $ a_n(\bf k)$ are the creation and annihilation operators, that obeys the anticommutation rules $\{ a_n^{\dagger}(\bk), a_n(\bk')\}=\delta_{nm}\delta(\bk-\bk')$. In this basis we have
\begin{eqnarray}
	{H}_0=\sum_{n\bk}\omega_n(\bk)a_n^\dagger(\bk)a_n(\bk)
\end{eqnarray}
with $\omega_n$ the energy of band $n$. The model of disorder potential arising from a serious randomly distributed impurities reads
\begin{eqnarray}
	V(\rv)=\sum_{\bf R}v(\rv -{\bf R})
\end{eqnarray}
where $v$ is the potential of a single impurity placed at the origin of the coordinate system, and $\bf R$ labels the coordinates. In the Bloch representation we can easily show 
\begin{eqnarray}
	V^{mm'}_{{\bf k}{\bf k}'}=v^{mm'}_{{\bf k}{\bf k}'}\sum_{\bf R}e^{-i({\bf k}-{\bf k}')\cdot {\bf R}}, \label{hamiltonianv}
\end{eqnarray} 
where we have used $v^{mm'}_{{\bf k}{\bf k}'}\equiv\langle m{\bf k}|V|m'{\bf k}'\rangle$ for the disorder potential matrix elements in a band eigenstates representation. We refer to the representation provided by the $|m \bk \rangle$ basis as the eigenstate
representation and write $|m \bk \rangle=e^{-i\bk \cdot \br} |u_{m\bk}\rangle$. In the limit of Born approximation, it is straightforward to show that $\langle V^{mm'}_{{\bf k}{\bf k}'} V^{m''m'''}_{{\bf k}'{\bf k}} \rangle=n_i  v^{mm'}_{{\bf k}{\bf k}'} v^{m''m'''}_{{\bf k}'{\bf k}}$~\cite{culcer2017interband}. Hence, in the Bloch representation we can write 
\begin{eqnarray}
	H_{imp}=\sum_{nm}\int d\bk d\bk' v_{\bk \bk '}^{mn}a_n^\dagger(\bk)a_m(\bk').
\end{eqnarray}
To proceed with Eq.~\ref{hamiltonian_p}, we need to identify the electric polarization ${\bf P}=e \rv$ in the basis of Bloch fields as
\begin{eqnarray}
	{\bf {P}}=e \sum_{nm{\bf k}{\bf k}'}\langle n{\bk}|\rv|m\bk'\rangle a_n^{\dagger}(\bk)a_m(\bk'),
\end{eqnarray}
where to proceed with above equation, one should derive an expression for the matrix element 
\begin{eqnarray}
	\langle n{\bk}|\rv|m\bk'\rangle  \equiv   \int d\rv \bpsi_n^\dagger(\bk,\rv) \rv  \bpsi_m(\bk', \rv). \label{forr} 
\end{eqnarray}
For Eq.\ref{forr}, Yet Bount~\cite{Blount1962} first identified these matrix elements as
\begin{eqnarray}\label{average_r}
	\langle n{\bk}|\rv|m\bk'\rangle&=&\delta_{nm}\left[\delta(\bk-\bk')\xi_{nn}(\bk)+i\partial_{\bk}\delta(\bk -\bk')\right]\nn\\
	&+&(1-\delta_{nm})\delta({\bk-\bk'})\xi_{nm}
\end{eqnarray}
containing the correlation $\xi_{mn}$ as
\begin{eqnarray}
	i\frac{\partial u_n(\bk,\rv)}{\partial \bk} =\sum_{m} u_m(\bk, \rv)\xi_{mn}
\end{eqnarray}
and for nondegenerate points in the Brillouin zone
\begin{eqnarray}
	\xi_{mn}=\frac{{\bf v}_{mn}(\bk)}{i \omega_{mn}(\bk)}.
\end{eqnarray}
The velocity matrix elements ${\bf v}_{mn}(\bk)$ is defined by 
\begin{eqnarray}
	\frac{1}{m}\int \psi_n^\dagger(\bk, {\bf x})  (-i\nabla)   \psi_m(\bk', {\bf x}) d{\bf x}= {\bf v}_{mn}(\bk) \delta(\bk -\bk')
\end{eqnarray}
and $m$ is the free electron mass. In order to make it convenient, we can use 
\begin{eqnarray}\label{epsilon}
	r_{nm}(\bk) &\equiv& \xi_{nm}(\bk) \qquad \qquad \textnormal{if} \quad n\neq m\nn\\
	&\equiv& 0   \quad\,\,\, \qquad \quad \qquad \textnormal{if}\quad n=m.
\end{eqnarray}
 According to Eq.~(\ref{average_r}), the polarization operator presented in Eq.~\ref{hamiltonian_p} can be separated into ``interband'' polarization $\hat{\bf P}_i$ for $(1-\delta_{nm})$ component and the ``intraband'' polarization $\hat{\bf P}_e$ polarization for $\delta_{nm}$ component
\begin{eqnarray}
	\hat{\bf P}&=&\hat{\bf P}_e+\hat{\bf P}_i\nn\\
	&=& e\sum_{nm\bk}r_{nm}(\bk)a_n^\dagger(\bk)a_n(\bk)\nn\\
	&+&ie\sum_{n\bk} a_n^\dagger(\bk)a_{n,b}(\bk)\label{p_inter}
\end{eqnarray}

In Eq.~\ref{p_inter}, the interaband polarization dependents on the covariant derivative 
\begin{eqnarray}
	a_{n,b}(\bk)\equiv\frac{\partial a_{n}(\bk)}{\partial{{k}^b}}-i(\xi_{mm}^b-\xi^b_{nn})a_{n}(\bk),
\end{eqnarray} 
where here and throughout, the Latin superscripts stand for Cartesian components $b=x,y,z$. The main result of this paper is to generate  quantum kinetic equation that accounts for disorder and electric field. Throughout this paper, we take disorder into account within the standard model of white-noise disorder potential, implicitly assuming therefore that disorder is weak. Although the extrinsic SOC can also be considered in the near future work. 
\section{Current density operator}\label{currentdensity1}
To establish some notations, in this part, we first consider the case when the external electric and magnetic fields are absent and then, in the next subsection, we will evaluate the Bloch equation in the more general when the electric fields are present. We start with a single-particle density matrix ${\rho}=e\bpsi^\dagger \psi$ described by the quantum Liouville equation
\begin{eqnarray}
	\frac{\partial {\rho} }{\partial t}=\frac{1}{i}[H,{\rho}],\label{liouville}
\end{eqnarray}
where $H$ is the total Hamiltonian Eq.~\ref{hamiltonian} of the complicated system containing the band Hamiltonian, disorder potential and a perturbation $H_E$ due to the external electric fields and all other interactions. To grasp proceed with the scattering process, it is convenient to divide the density matrix into two parts
\begin{eqnarray}
	\rho=\langle \rho \rangle + {g},
\end{eqnarray} 
where $\langle \rho \rangle$ denotes the average over impurity configuration and ${g}$ is the reminder where its average over impurity configuration vanishes $\langle {g}\rangle$. To keep the discussion as simple as possible, we confine first to the case when the electric field is absence. The extension results to $H_E\neq0$ is straightforward. In Ref.~\cite{culcer2017interband, Culcer2017}, a general quantum kinetic theory of linear response to an external electric field applied to solids with arbitrarily complicated band structures including the interband and intraband coherence response and the formalism was based on density-matrix equations of motion limited to the Born approximation treatment of disorder. To present a consistent analysis we first recall the key steps of their derivations. The starting point is the quantum Liouville Eq.~\ref{liouville} expands to the expressions for $\langle \rho \rangle$ and $g$ as follows 
\begin{align}
	{}&\frac{d\langle \rho \rangle}{dt}+{i}[H_0,\langle \rho \rangle]+{i}\langle [v,g]\rangle=0,\label{g0}\\
	{}&\frac{d {g}}{dt}+{i}[H_0,g]+{i}[v,\langle \rho \rangle]=-{i}[v,\hat{g}]+{i}\langle [v,{g}]\rangle.\label{g1}
\end{align}
In Eq.~\ref{g1} we can ignore both terms on the right hand side since they are beyond the Born approximation, we can ignore them. Then from Eqs.~\ref{g1} and Eq.~\ref{g0}, we can obtain~\cite{vasko2006quantum}
\begin{eqnarray}\label{g2}
	g(t)={-i} \int_{0}^{\infty} dt' [e^{-iH_0t'}Ve^{iH_0t'},\langle \rho(t)\rangle],
\end{eqnarray}
where we have used the disorder-free expression for the time evolution of the density matrix as follows
\begin{eqnarray}
	\langle\rho(t-t')\rangle=e^{-iH_0t'}\langle \rho (t) \rangle e^{iH_0t'}.
\end{eqnarray}
Hence by inserting Eq.~\ref{g2} into Eq.~\ref{g0} we can obtain the full kinetic equation as
\begin{eqnarray}\label{kinetic}
	\frac{d\langle \hrho \rangle}{dt}+\frac{i}{\hbar}[H_0,\langle \hrho \rangle]+J(\langle\hrho\rangle)=0,
\end{eqnarray}
where the electron-impurity collision integral is given by 
\begin{equation}
	J(\langle\hrho\rangle)=\int_{0}^{\infty} dt'\langle [V,[e^{-iH_0t'}v e^{iH_0t'},\langle \rho (t) \rangle ]]\rangle.
\end{equation}
In this equation we notice that the dependence of $\langle \rho(t) \rangle$ to the disorder potential is no longer explicitly and at this moment we can take into account the translational symmetry recovered after impurity averaging by working in momentum
space where $J_{\bk}(\langle\rho\rangle)$ is diagonal.  In the Bloch representation the collision integral at the momentum $\bk$ is given by 
\begin{eqnarray}\label{collision}
	J_{\bk}(\langle\rho\rangle)&=&\int_{0}^{\infty} d{t'} \sum  \\ 
	&& \bigl\{ \langle U^{mm'}_{\bk \bk'}U^{m'm''}_{\bk' \bk} \rangle e^{-i(\omega^{m'}_{\bk'}-\omega^{m''}_{\bk})t'} \langle \rho\rangle^{m''m'''}_{\bk}\nn\\
	&-&\langle U^{mm'}_{\bk \bk'}U^{m''m'''}_{\bk' \bk} \rangle \langle \rho\rangle^{m'm''}_{\bk'}e^{-i(\omega^{m''}_{\bk'}-\omega^{m'''}_{\bk})t'}\nn\\
	&-&\langle U^{mm'}_{\bk \bk'}U^{m''m'''}_{\bk' \bk} \rangle e^{-i(\omega^{m}_{\bk}-\omega^{m'}_{\bk'})t'} \langle \rho\rangle^{m'm''}_{\bk'}\nn\\
	&+&\langle U^{m'm''}_{\bk \bk'}U^{m''m'''}_{\bk' \bk} \rangle \langle \rho\rangle^{mm'}_{\bk}e^{-i(\omega^{m'}_{\bk}-\omega^{m''}_{\bk'})t'}\bigr\}.\nn 
\end{eqnarray}
The integral over $t'$ can be obtained by regularizing the time integral by inserting a convergence factor $e^{\eta t'}$ as follows
\begin{eqnarray}
	\int_{0}^{\infty} &dt'&e^{-i(\omega^{m}_{\bk}-\omega^{m'}_{\bk'}-i\eta)t'}\label{integral}\\   
	&\equiv&\mathcal{P}[\frac{1}{i(\omega^{m}_{\bk}-\omega^{m'}_{\bk'})}]
	+{\pi}\delta(\omega^{m}_{\bk}-\omega^{m'}_{\bk'})\nn
\end{eqnarray}
where $\delta$ part give us the conservation law, whereas the principle part presents the disorder-induced level repulsion. Now we are able to back to Eq.~\ref{kinetic} and solve the kinetic equation by decomposing the disorder averaged part of the density matrix into two parts, $\langle \rho \rangle=n+S$, where $n$ is diagonal in band index and $S$ is off-diagonal in band index. By using Eq.~(\ref{integral}), the diagonal part for the energy conservation term reads 
\begin{align}
	 [J_d(\rho)]^{mm}_{\bk}&=2\pi n_i\sum_{m'\bk'}u^{mm'}_{\bk\bk'}u^{m'm}_{\bk'\bk}(\rho^{mm}_{\bk}-\rho^{m'm'}_{\bk'})\delta(\omega^m_{\bk}-\omega^{m'}_{\bk'})\nn \label{jdmm}\\ \\
	[J_d(\rho)]^{mm''}_{\bk}&={\pi n_i}\sum_{m'\bk'}u^{mm'}_{\bk\bk'}u^{m'm''}_{\bk'\bk}\\
	{}&\quad\quad \left[(\rho^{mm}_{\bk}-\rho^{m'm'}_{\bk'})\delta(\omega^m_{\bk}-\omega^{m'}_{\bk'})\right.\nn \\
	&\quad \qquad\quad\left.+(\rho^{m''m''}_{\bk}-\rho^{m'm'}_{\bk'})\delta(\omega^{m''}_{\bk}-\omega^{m'}_{\bk'})\right]\nn \label{spincurrent}
\end{align}
where Eq.~\ref{jdmm} presents the Fermi's golden role when $m=m''$ and Eq.~\ref{spincurrent} is raised when for $m\neq m''$. For principle part of the collision integral term we get 
\begin{eqnarray}
	[J_{pp}(\rho)]^{mm''}_{\bk}&=&{n_i}\sum u^{mm'}_{\bk\bk'}u^{m'm''}_{\bk'\bk}\\
	&&\left[ (\rho^{m''m''}_{\bk}-\rho^{m'm'}_{\bk'})\mathcal{P}(\frac{1}{\omega^{m'}_{\bk'}-\omega^{m''}_{\bk}})\right. \nn \\
		&+&\left. (\rho^{mm}_{\bk}-\rho^{m'm'}_{\bk'})\frac{1}{\omega^{m}_{\bk}-\omega^{m'}_{\bk'}}\right]\nn 
\end{eqnarray}
Notice that the diagonal elements $m=m''$ vanishes for the principle part. In our analysis the principle terms are not considered and only we take into account delta function of Eq.~\ref{integral} since the real part has been absorbed in just a shift of the Fermi level and the $\delta$-function part becomes important near points in momentum space where different bands touch.

\subsection{Applying the electric field into the Kinetic equation}

In this subsection, we add the external electric field to the system. As we mention before, when $E\neq0$, the term ${H}_{\bf E}$  adds to the total Hamiltonian. We assume that the electric field is small like disorder term, and with this assumption, we are able to use the perturbation theory. According to Eq.~\ref{liouville}, the quantum Liouville equation has the form 
\begin{align}
	{}&\frac{d \rho }{dt}+{i}[H_0,\rho]+J(\rho)=-{i}[H_{\bf E},\rho]
\end{align}
Now we can solve the quantum kinetic perturbatively by applying a double expansion on the impurity concentration $n_i$. Note that in this present work we confine the results in the Born approximation and we neglect higher order in impurities like the skew scattering effect and side-jump effect. To do so, we derive the density matrix first to linear order to the electric field and then we extend our results to second order. In linear response the density matrix $\langle \rho \rangle$ can decompose into the equilibrium $\langle \rho_0 \rangle $, and the correlation to the equilibrium density matrix to first order in the electric field $\langle \rho_{E} \rangle$. Hence $\langle \rho \rangle=\langle \rho_0 \rangle+ \langle \rho_E \rangle$. By linearizing the quantum kinetic equation, we obtain 
\begin{eqnarray}
 \frac{d \langle \rho_{E} \rangle }{dt}+{i}[H_0,\langle \rho_{E} \rangle]+J(\langle \rho_{E} \rangle)=-{i}[H_E,\langle \rho_{0} \rangle] -{i} [H_E,\langle \rho_{E} \rangle]\nn \\ 
 \label{linear_kinetic}
\end{eqnarray}
where $ \langle \rho_E \rangle=\lambda^1 \langle \rho^1 \rangle+\lambda^2 \langle \rho^2 \rangle+ \cdots$ with expanding in the external electric field. Notice that in the  response theory, we can ignore the latest term on right hand side of Eq.~\ref{linear_kinetic} because it contributes one order beyond of the electric field. Now we can start with driving term on the right hand side of above equation by using the Bloch eigenstate representation
\begin{eqnarray}
	{-i} [H_E, \langle \rho_0 \rangle]&=&{-e}\sum_{b} E^b(t) \langle  \rho_{0} \rangle_{mn,b}\\
	&+& {ie} \sum_{bp} E^b(t) \left( r^b_{mp}\langle \rho_{0} \rangle_{pn}  -\langle \rho_{0}\rangle_{mp} r^b_{pn}\right).\nn
\end{eqnarray}
In the Bloch eigenstate representation, the ground state representation for the density matrix was given by 
\begin{eqnarray}
	\langle \rho_{0}\rangle_{mn}= f_0(\omega^{m}_{\bk})\delta_{mn}\equiv n^{mm}_{0\bk}, \label{ground} 
\end{eqnarray}
where $f_0(\omega_{m}^{\bk})$ presents the Fermi-Dirac distribution function evaluated at energy  $\omega_{m}^{\bk}$. Therefore by using Eq.~(\ref{ground}) we can immediately derive 
\begin{eqnarray}
	{-i} [H_E,\langle \rho_{0} \rangle_{nm}]&=& {-e {\bf E}}\cdot \left[\delta_{mn}\frac{\partial f_0(\omega^{n}_{\bk})}{\partial{\bk}}\right.\label{hEp0}\\
	&&\qquad-i\left. \sum_{p}{\bf r}_{nm}\left(f_0(\omega^{n}_{\bk})-f_0(\omega^{m}_{\bk})\right)\right]\nn 
\end{eqnarray} 
The later term in square brackets denotes off-diagonal electric response when $m\neq m'$, and is responsible for the interband coherent contributions to the electric response of solids. By using Eq.~\ref{hEp0} and limiting to the Born approximation for the collision integral, we can derive the density matrix for the diagonal terms and then for off-diagonal part is straight forward. Hence by using Eq.~\ref{linear_kinetic} for the diagonal part of the density matrix $m=m'$, we have  
\begin{eqnarray}
	i\frac{d\langle \rho^1 \rangle_{mm}}{dt}-[H_0,\langle \rho^1 \rangle]_{mm}-\langle[v, \rho^1] \rangle_{mm}=[H_E,\langle \rho^0 \rangle]_{mm}\nn \\
	\label{diagonalp}
\end{eqnarray}
where the second term on the left hand side of above equation is purely off-diagonal in the band index and it can be neglected for diagonal part of the density matrix. The later term can replace by Eq.~\ref{hEp0} where just the first term on that equation is diagonal in matrix space. 
The collision term on the later term of Eq.~\ref{diagonalp}, can be seen as a matrix operator acting on the density matrix. For the diagonal response we can define it according to 
\begin{eqnarray}\label{collision}
	{i}\langle[v, \rho^1] \rangle_{mm}\equiv J_d(\rho^1)^{mm}_{\bk}=n_i\sum_{m'\bk'}\mathcal{K}^{mm'}_{\bk \bk'}\rho^{1m'm'}_{ \bk'}
\end{eqnarray}
where $\mathcal{K}$ acts on the density matrix as a vector with components labeled by band and wave vector. In comparison with Eq.~\ref{jdmm}, we have
\begin{eqnarray}
	\mathcal{K}^{mm'}_{\bk\bk'}&= &{2\pi n_i}\left[\delta_{\bk\bk'}\delta_{mm'}\sum_{m'',\bk''}u^{mm''}_{\bk\bk''}u^{m''m}_{\bk''\bk}\delta(\omega^{m}_{\bk}-\omega^{m'}_{\bk'})\right. \nn  \\
	&&\qquad\qquad \quad \quad -\left. u^{mm'}_{\bk\bk'}u^{m'm}_{\bk'\bk}\delta(\omega^{m}_{\bk}-\omega^{m'}_{\bk'})\right]\label{relaxationf}
\end{eqnarray}
where  for the simplicity of the calculations from now we consider $\langle \hrho^1\rangle=\hrho^1$.  
Ansatz for the diagonal matrix in longer time limit,  
\begin{equation}
	\rho_{{\bf k}, {\bf k}}^{1mm}=\sum_{b\beta}\widetilde{\rho}^{1, b\beta}_{m{\bf k}, m{\bf k}}E^b_\beta e^{-i \omega_{\beta}t},\label{rho1}
\end{equation}
where monochromatic electric field is defined by 
\begin{equation}
	E^b=\sum_{\beta}E_{\beta}^{b}e^{-i\omega_{\beta}t}.
\end{equation}
By taking the derivation of Eq. \ref{rho1}, and inserting Eq.~\ref{collision} and Eq.~\ref{hEp0} inside Eq. \ref{diagonalp}, we can easily show that 
\begin{equation}
	(-i\omega_\beta)\hrho^{1,b\beta}_{m{\bf k}, m{\bf k}}+n_i\sum_{m'\bk'}\mathcal{K}_{\bk \bk'}^{mm'}\hrho^{1,b\beta}_{m'{\bf k}', m'{\bf k}'}={e}\frac{\partial{f({\omega^m_{\bk}})}}{\partial{k}^b}
\end{equation}
By inserting Eq.~\ref{rho1} to Eq.~\ref{diagonalp}, we read
\begin{equation}
	\hrho_{{\bf k}, {\bf k}}^{1mm}={e}\sum_{b\beta}\sum_{m'\bk'} (\Gamma^{mm'}_{\bk \bk'})^{-1}\frac{\partial{f({\omega_{\bk}^m})}}{\partial{k'}^b} E^b_\beta e^{-i \omega_{\beta}t}\label{currentdensity}
\end{equation}
where
\begin{align}
{}&	\Gamma^{mm'}_{\bk \bk'}=-i\omega_{\beta}\delta({\bk-\bk'})\delta_{mm'}+n_i\mathcal{K}^{mm'}_{\bk \bk'},\\
{}&	\frac{\partial{f({\omega_{\bk}^m})}}{\partial{k}^b}= \hat{k}_b v^m_k \delta(f({\omega_{\bk}^m})-\omega_F), \\
{}&	\hat{\bv}_{\bk}=\hat{k}\partial_{\bk}-\frac{\hat{\psi}}{k}\partial_{\psi}, \qquad \text{with \ \ }\hat{\psi}=(-\sin\psi,\cos\psi),
\end{align}
and $\omega_F$ is the Fermi level energy. It is clear that in the absence of the disorder potential, the diagonal term in density matrix is zero and the there is only a contribution to $\hrho_{{\bf k}, {\bf k}}^{1mm'}$ from the interband coupling when $m\neq m'$. This is in agreement with the results reported in Ref.~\cite{sipe2000, sipe1993}. 
The collision term in Eq.~\ref{collision} can be seen as a matrix operator acting on the density matrix. By solving for $\rho_{\bk}^{(-1),mm}$ to leading order in $n_i$ according to Eq.~\ref{diagonalp}, we have
\begin{eqnarray}\label{p1diagonal}
	\rho_{\bk}^{(-1),mm}=n_i^{-1}e {\bf E}\cdot\sum_{m\bk'}({\mathcal{K}^{mm'}_{\bk\bk'}})^{-1}\cdot{\bf v}^{m'}_{\bk'}\frac{\partial f_0(\omega^{m'}_{\bk'})}{\partial \omega^{m'}_{\bk'}}
\end{eqnarray} 
where we have used ${\bf v}^{m}_{\bk}=\partial\omega_{\bk}^{m}/\partial{\bk}$. In the simplest case of metals and insulators we have 
\begin{eqnarray}
	\frac{\partial f_0(\omega_{\bk}^m)}{\partial\omega_{\bk}^m}&=&-\delta(\omega_{\bk}^m-\omega_F) \quad \textnormal{for metals}\\
	&=&0 \ \  \qquad \qquad\qquad \textnormal{for insulators}
\end{eqnarray}
It is clear that the band-diagonal part of the density matrix Eq.~\ref{p1diagonal}  is a contribution to balance the intraband driving term and scattering on the Fermi surface and in the limit of very weak disorder scattering diverges. Now we can derive the off-diagonal terms in the density matrix. Since $ [H_0, \langle \rho_{0} \rangle]$ is purely off-diagonal, we can conclude that the diagonal part on the density matrix is never divergence even in the absence of disorder potential. Hence the leading response is independent of disorder strength $n_i$ as
\begin{eqnarray}
	\frac{\partial S^{1,mm'}}{\partial{\bk}} +{i}[H_{0\bk},S^{1}_{\bk}]^{mm'}=D^{mm'}_{\bk \bk'}+D^{'mm'}_{\bk \bk'}\label{offdiagonal}
\end{eqnarray}
where the first term on the right hand side is intrinsic off-diagonal part shown in later term of Eq.~(\ref{hEp0})
\begin{eqnarray}
	D_{Ek}^{mm'}= {ie{\bf E}}\cdot\sum_{\bk}{\bf r}_{mm'}\left(f_0(\omega_{\bk}^m)-f_0(\omega_{\bk}^{m'})\right)
\end{eqnarray}
and $D^{'mm'}_{\bk \bk'}$ is responsible for disorder scattering $D^{'mm'}_{\bk \bk'}=-J_{d}[\rho_{\bf k}^{1}]_{mm'}$ which is non zero at Fermi surface. By using the matrix element $J_{d}[\rho_{\bf k}^{1}]_{mm'} $ from Eq.~\ref{jdmm},  $D^{'mm'}_{\bk \bk'}$ takes the form
\begin{eqnarray}
	D^{'mm'}_{\bk \bk'}&=&{\pi n_i}\sum_{m'\bk'}u^{mm'}_{\bk\bk'}u^{m'm''}_{\bk'\bk} \\
	&&\left[(\rho^{-1,mm}_{\bk}-\rho^{-1,m'm'}_{\bk'})\delta(\omega^m_{\bk}-\omega^{m'}_{\bk'})\right. \nn \\
	&+&\left. (\rho^{-1,m''m''}_{\bk}-\rho^{-1,m'm'}_{\bk'})\delta(\omega^{m''}_{\bk}-\omega^{m'}_{\bk'})\right].\nn
\end{eqnarray}
By solving Eq.~\ref{offdiagonal}, we have 
\begin{eqnarray}
	S^{1,mm'}_{\bk}=\int_{0}^{\infty} dt'e^{-iH_0(\bk)t'}(D^{mm'}_{\bk \bk'}+D^{'mm'}_{\bk \bk'})e^{iH_0(\bk)t'}\nn \\
\end{eqnarray}
After integrating over time, the off-diagonal density matrix in linear order of the electric field yields  
\begin{eqnarray}
	S^{1,mm'}_{\bk}&=&-i \mathcal{P}\left(\frac{D^{mm'}_{\bk \bk'}+D^{'mm'}_{\bk \bk'}}{\omega_{\bk}^m-\omega_{\bk'}^{m'}}\right) \\
	&+&\pi \left(D^{mm'}_{\bk \bk'}+D^{'mm'}_{\bk \bk'}\right)\delta(\omega_{\bk}^m-\omega_{\bk'}^{m'})\nn 
\end{eqnarray}
As a result, we can write the linear response of the density matrix in the present of disorder potential accordingly
\begin{equation}
	\rho^{1, m m'}_{\bk \bk'}=\rho^{1, mm'}_{\bk}\delta_{mm'}+\rho^{1, m m'}_{\bk \bk'},\label{density}
\end{equation}
where the first term describes the diagonal elements of the density matrix to the first order in the electric field and finite frequency in long term limit derived by Eq.~\ref{currentdensity}, and the second term is off-diagonal density matrix to finite frequency when $m\neq m'$ is
\begin{eqnarray}\label{off-diagonal1}
	\rho^{1, m m'}_{\bk \bk}=-i\sum_{b\beta}\frac{ \widetilde{D}^{b\beta, mm'}_{\bk \bk}+\widetilde{I}^{b\beta, mm'}_{\bk \bk}}{(\omega^m_{\bk}-\omega^{m'}_{\bk})-\omega_\beta}E^b_\beta e^{-i\omega_{\beta}t}
\end{eqnarray}   
with
\begin{eqnarray}
	\widetilde{D}^{b\beta, mm'}_{\bk \bk}&=&{i e}r^{b}_{m m'}(\bk) \big(f(\omega^m_{\bk})-f(\omega^{m'}_{\bk})\big)\label{off-diagonal11}\\
	\widetilde{I}^{b\beta, mm'}_{\bk \bk'}&=&{\pi n_i}\sum_{m'',\bk''}u^{mm'}_{\bk\bk'}u^{m'm''}_{\bk'\bk}\label{off-diagonal12}\\
	&&\big[(\widetilde{\rho}_{\bk}^{1b\beta, mm}-\widetilde{\rho}_{\bk'}^{1b\beta, m'm'})\delta(\omega^m_{\bk}-\omega^{m'}_{\bk'})\nn \\
	&+& (\widetilde{\rho}_{\bk}^{1b\beta, m''m''}-\widetilde{\rho}_{\bk'}^{1b\beta,m'm'})\delta(\omega^{m''}_{\bk}-\omega^{m'}_{\bk'})\big]\nonumber
\end{eqnarray}
The off-diagonal matrix presenting in Eq.~\ref{off-diagonal1} consists of two main parts: First part shown in Eq.~\ref{off-diagonal11} is independent of disorder character and is an intrinsic band-structure property involving the entire Fermi sea~\cite{sipe1993,sipe2000}. Second one shown in Eq.~\ref{off-diagonal12} is an extrinsic contribution containing both the band off-diagonal and the band-diagonal parts of the density matrix. Obviously it is finite in the weak disorder limit according to Eq.~\ref{p1diagonal} and originates from the collision kernel acting on the leading Fermi surface response.

\subsection{Second-order Electric Response}
In this subsection, we extend our results to the second order electric response for the density matrix. To do so, we start from the equation of the motion for ${\rho}^2$ accordingly
\begin{eqnarray}
	\frac{d \langle \rho^{2} \rangle }{dt}+{i}[H_0,\langle \rho^{2} \rangle]&+&J(\langle \rho^{2} \rangle)=\\
	&&- {i}[H_E,\langle \rho^{1} \rangle]-{i}[H_E,\langle \rho^{2} \rangle],\nn \label{secondrho}
\end{eqnarray}
where the last term on the right hand side is beyond second order and is negligible. To derive the diagonal term, also we can neglect the second term on Eq.~\ref{secondrho} since it is purely off-diagonal in matrix space.  As we derived in the linear response, we must derive all the terms presented in Eq.~\ref{secondrho}. Hence the electric dependency is derived by 
\begin{eqnarray}
	{-i} \langle mk|[H_E, \rho^1 ]|m'k'\rangle&=& {-e {\bf E}}\cdot \left[\delta_{mm'}\frac{\partial \langle \rho^1 \rangle^{mm'}_{\bk}}{\partial{\bk}}\right.\nn \\
	&&\quad-\left. i[\mathcal{R}_{\bk},\langle \rho^1\rangle_{\bk}]^{mm'}\right],   
\end{eqnarray} 
where $\mathcal{R}^{mm'}_\bk=\langle u^m_{\bk}|i\frac{\partial u^{m'}_{\bk} }{\partial{\bk}}\rangle$ and the first term is derived by inserting the diagonal term of density matrix shown in Eq.~\ref{density} and take the derivation of the momentum to the second order. Hence with doing some calculations the diagonal term to second order response of the density matrix yields  
\begin{eqnarray}
	\rho_{m\bk m\bk}^{(2), b\beta c \gamma}&=&{e}\sum_{m'\bk'} ({\Gamma^{mm'}_{\bk\bk'}})^{-1} \big[\partial_{k_c}\sum_{m''\bk''}\big(({\Gamma^{m'm''}_{\bk' \bk''}})^{-1}\frac{\partial f (\omega^{m''}_{\bk''})}{\partial_{\bk''}}\big)\nn\\
	&-&{i}{e}\sum_{m'\bk'} {(\Gamma^{mm'}_{\bk \bk'})}^{-1} \sum_{n}\big(r^c_{mn}B^b_{nm}-B^b_{mn}r^c_{nm}\nn\\
	&&\quad\qquad\qquad\qquad\qquad+r^c_{mn}\widetilde{I}^{'b\beta}_{nm}-\widetilde{I}^{'b\beta}_{mn}r^c_{nm}\big)\big]\label{diagonalp2}
\end{eqnarray}
where 
\begin{eqnarray}
	B^b_{nm}=\frac{ef_{mn}r^b_{mn}}{\omega^{mn}-\omega_\beta},\quad \widetilde{I}^{'b\beta}_{nm}=\frac{\widetilde{I}^{b\beta}_{nm}}{\omega^{mn}-\omega_\beta},
\end{eqnarray}
including $f_{nm}=f(\omega^n_{\bk})-f(\omega^m_{\bk})$ and $\omega^{mn}=\omega^m_{\bk}-\omega^n_{\bk}$. For the off-diagonal terms, we can start from Eq.~\ref{secondrho} with a point that in this part we have to consider the second term on the left hand side since this term is purely off-diagonal and need to be considered. Hence we have 
\begin{eqnarray}
	{i}[H_0,\langle \rho^{2} \rangle]^{mm'}&=&i\omega^{mm'}_{\bk \bk'}\rho^{2,mm'}_{\bk \bk'}
\end{eqnarray}
It is obviously that the above equation is purely of diagonal. The next term on Eq.~\ref{secondrho} is written as 
\begin{eqnarray}
	 &&-{i}[H_E,\langle \rho^{(1)} \rangle]^{mm''}={e}\sum_{b\beta}\sum_{c\gamma}\big[\partial_{k_c}\big(B^b_{mm''} +\widetilde{I}^{'b\beta}_{mm''}\big)\nn \\ \nn
	&&\qquad \qquad\quad-i\big(r^c_{mm''}-r^c_{m''m}\big)\sum_{m'\bk'}(\Gamma^{mm'}_{\bk \bk'})^{-1}\frac{\partial f(\omega^{m'}_{\bk'})}{\partial{k^{'b}}}\\ \nn 
	&&\qquad\quad \qquad+\sum_{m'} \bigg( r^c_{mm'} B^b_{m'm''}-B^b_{mm'}r^c_{m'm''}\nn \\
	&&\qquad\qquad \qquad\qquad\qquad+i(r^c_{mm'} \widetilde{I}^{'b\beta}_{m'm''}-\widetilde{I}^{'b\beta}_{mm'}r^c_{m'm''} ) \bigg)\nn \\
	&&\quad\qquad \qquad-\big(\omega^c_{mm}-\omega^c_{m''m''}\big)\big(iB^b_{mm''}+\widetilde{I}^{'b\beta}_{mm''}\big) \big]\nn \\
	&&\qquad \qquad\qquad\qquad\times  E^b_\beta E^c_\gamma e^{-i(\omega_{\beta}+\omega_{\gamma})t} \nn  \\
	&&\qquad \qquad\quad\quad\qquad\qquad\equiv \widetilde{\eta}^{b\beta c \gamma }_{mm''} E^b_\beta E^c_\gamma e^{-i(\omega_{\beta}+\omega_{\gamma})t} \label{ep2}
\end{eqnarray}
In above equation, when $v_{imp}=0$, the latter term vanishes and we have 
\begin{eqnarray}
	-{i} [H_E,\langle \rho^{(1)} \rangle]^{mm''}&=&{e} \sum \big[ r^c_{mm'} B^b_{m'm''} -B^b_{mm'}r^c_{m'm''}  \big]\nn \\
	&&\qquad\qquad \times E^b_\beta E^c_\gamma e^{-i(\omega_{\beta}+\omega_{\gamma})t}
\end{eqnarray}
This is in agreement with Refs.~\cite{sipe1993, sipe2000} when the second order of the electric response is recorded in clean limit. Now the collision integral is founded as 
\begin{eqnarray}
	J_{imp}(\rho^{2})&=&{\pi n_i} \sum_{m'\bk'}u_{\bk\bk'}^{mm'}u_{\bk'\bk}^{m'm}\nn \\
	&\times& \big[ (\rho^{2,b\beta c\gamma}_{m\bk m\bk}-\rho^{2,b\beta c\gamma}_{m'\bk' m'\bk'})\delta(\omega^{m}_{\bk}-\omega^{m'}_{\bk'})\nn  \\ 
	&+& (\rho^{2,b\beta c\gamma}_{m''\bk m''\bk}-\rho^{2,b\beta c\gamma}_{m'\bk' m'\bk'})\delta(\omega^{m''}_{\bk}-\omega^{m'}_{\bk'})\big]\nn \\
	&\times& E^b_\beta E^c_\gamma e^{-i(\omega_{\beta}+\omega_{\gamma})t} \nn  \\
	&\equiv& \mathcal{I}^{b \beta c \gamma}_{m \bk m'\bk'}E^b_\beta E^c_\gamma e^{-i(\omega_{\beta}+\omega_{\gamma})t}\label{jp2}
\end{eqnarray}
Now we are able to back to Eq.~\ref{secondrho}. By inserting Eqs.~\ref{jp2} and \ref{ep2} to Eq.~\ref{secondrho}, we have 
\begin{eqnarray}
	\rho^{2mm''}_{\bk} =-\sum_{b \beta}\sum_{c\gamma} \frac{\widetilde{\eta}^{b\beta c \gamma }_{mm''}+\widetilde{I}^{b\beta c \gamma}_{mm''}}{(\omega_{\bk}^m-\omega_{\bk}^{m''})-\omega_{\Sigma}} E^b_\beta E^c_\gamma e^{-i\omega_{\Sigma}t}\label{rho2} 
\end{eqnarray}
with $\omega_{\Sigma}=\omega_\beta+\omega_{\gamma}$. In order to compare our results with previous results derived by Refs.~\cite{sipe1993, sipe2000} when $V_{imp}=0$, we have 
\begin{eqnarray}
	\widetilde{\eta}^{b\beta c \gamma }_{mm''}={e}\big(B^{b}_{mm'',c}+\sum_{m'} (r^{c}_{mm'}B^{b}_{m'm''}-B^{b}_{mm'}r^c_{m'm''}) \big)\nn \\ \label{eta}
\end{eqnarray}
where 
\begin{eqnarray}
	B^{b}_{mm'',c}=\frac{\partial B^b_{mm''}}{\partial{k^c}}-i(\omega^{mm}_c(\bk)-\omega^{m''m''}_c(\bk)).
\end{eqnarray}
By inserting Eq.~\ref{eta} in Eq.~\ref{rho2}, when $V_{imp}=0$ we integrate to find 
\begin{eqnarray}
	\rho^{2mm'' }_{\bk} &=&\frac{-i e}{(\omega^m_{\bk}-\omega^{m''}_{\bk})-\omega_{\Sigma}} \sum_{b \beta}\sum_{c\gamma}\big(B^{b}_{mm'',c} \\ &&+\sum_{m'}(r^{c}_{mm'}B^{b}_{m'm''}-B_{mm'}^{b}r^{c}_{m'm''}) \big) E^b_\beta E^c_\gamma e^{-i\omega_{\Sigma}t}\nn
\end{eqnarray}
This is exactly in agreement with the previous results reported in~\cite{sipe1993, sipe2000}. Hence, density matrix to the second order response reads as 
\begin{equation}
	\rho^{2 m m'}_{\bk \bk'}=\rho^{2 mm'}_{\bk}\delta_{mm'}+\rho^{2 m m'}_{\bk \bk'},\label{densityp2}
\end{equation}
where the first part is the diagonal part derived in Eq.~\ref{diagonalp2} and the second one is off-diagonal reported explicitly on Eq.~\ref{rho2}. 

\section{Disordered Rashba model for a nonmagnetic semiconductor}\label{disorderedRashba}
The Hamiltonian model for a 2DEG in the presence of Rashba SOC (RSOC) reads 
\begin{eqnarray}
	H=\frac{k^2}{2m}+\alpha (k_y\sigma^x-k_x\sigma^y)+V_{imp}(\bf r)
\end{eqnarray}
with $m^{\star}$ the electron effective mass, $\alpha$ the spin-orbit strength and $\sigma^i$ the Pauli matrices $i=x,y,z$. For Hamiltonian the eigenvectors are 
\begin{eqnarray}
	|mk \rangle \equiv |\pm \bk \rangle = |u_{\bk}^{\pm}\rangle=\frac{1}{\sqrt{2}}
	\begin{pmatrix}
		e^{-i \theta}\\
		\mp i
	\end{pmatrix}\label{upm}
\end{eqnarray}
where $\theta$ is the polar angle of the wave vector. The energy splitting for each band read as
\begin{equation}
	\omega^{\pm}_{\bk}=\frac{k^2}{2m^{\star}}\pm \alpha k.\label{energyband}
\end{equation} 
where $m^{\star}$ is the effective mass of electron. 
In order to derive the relaxation time for interband and intraband, we could consider two different limits, short range impurity potential and long range impurity potential. In the short range of impurity we assume that the $V^{mm'}_{\bk \bk'}$ does not depend on ${\bf q}=\bk-\bk'$ and one can replace $V^{mm'}_{\bk \bk'}$ to $v_0$. On other side, when we assumed the impurity is large in its potential, according to Eq.~\ref{upm} we can derive  
\begin{eqnarray}
	\langle V^{\pm\mp}_{\bk\bk'}V^{\mp\pm}_{\bk'\bk}\rangle&=&\frac{{v}^2_0}{2}(1-cos\gamma)\\
	\langle V^{\pm \pm}_{\bk\bk'}V^{\mp \mp}_{\bk'\bk}\rangle&=&\frac{{v}^2_0}{2}(1+cos\gamma)\\
	\langle V^{++}_{\bk \bk'}V^{+-}_{\bk'\bk } \rangle&=&\frac{{v}^2_0}{2}sin\gamma\\
	\langle V^{+-}_{\bk \bk'}V^{--}_{\bk'\bk } \rangle &=&\frac{-{V}^2_0}{2}sin\gamma
\end{eqnarray}
In order to derive the matrix of relaxation time, the starting point is getting back to Eq.~\ref{relaxationf} for $\mathcal{K}^{mm'}_{\bk\bk'}$.
In short range impurity potential, as the scattering amplitude is constant and we can replace $V_{\bk \bk'}^{mm'}$ to ${v}_0$ as
\begin{eqnarray}
	\frac{1}{\tau_{mm'}}=\mathcal{K}^{mm'}_{\bk\bk'}={2\pi n_i{v}^2_0} \sum_{\bk'}(1-cos\gamma)\delta(\omega^{m}_{\bk}-\omega^{m'}_{\bk'}) 
\end{eqnarray}
that with integrating over momentum we read
\begin{eqnarray}
	\frac{1}{\tau_{mm'}}={n_i{v}^2_0}\int k' dk' \delta(\omega^{m}_{\bk}-\omega^{m'}_{\bk'}). \label{tmm}
\end{eqnarray}
In this case for the relaxation matrix, only term survived when $m=m'$ in diagonal terms and for off-diagonal terms $m\neq m'$ since $\gamma$ goes to zero, these terms vanishes. By considering the long-range impurity potential, by solving Eq.~\ref{relaxationf} to the case when the scattering amplitude dependents on $\gamma$ we can show that
\begin{eqnarray}
	\frac{1}{\tau_{mm'}}&=&{2\pi n_i{v}^2_0}\big(\frac{1}{4} 		 \sum_{\bk'}\delta(\omega^{m}_{\bk}-\omega^{m'}_{\bk'})\label{relaxationmm}\\
	&&\qquad \qquad \qquad-\frac{1}{4}\sum_{\bk'}cos(2\gamma)\delta(\omega^{m}_{\bk}-\omega^{m'}_{\bk'})  \big) \nn  
\end{eqnarray}
By taking the integration over the momentum, Eq.~\ref{relaxationmm} reads as
\begin{eqnarray}
	\frac{1}{\tau_{\pm \pm}} \equiv \frac{1}{\tau_{\pm}}&=&{n_i{v}^2_0} \big(\frac{1}{4} \int k' dk' \delta(\omega^{\pm}_{\bk}-\omega^{\pm}_{\bk'})\nn \\ 
	&&\qquad \qquad+ \frac{3}{4}\int k' dk' \delta(\omega^{\pm}_{\bk}-\omega^{\mp}_{\bk'}) \big)
\end{eqnarray}
In the long impurity regime, our equation shows that when $\gamma$ is taken into account, the interband relaxation time $m=m'$ is three times smaller than the intraband relaxation time $m\neq m'$.  
In order make our method as simple as possible, we confine the system to the short-range impurity potential. 
From now on, we limit the system to the short-range impurities and as considered before in this regime, only one relaxation time for each band is considered. More precisely, in the presence of RSOC, the SO splitting and the disorder broadening are much smaller than the Fermi energy $\omega_F$
\begin{eqnarray}
	\omega_F\gg \frac{1}{\tau_0}, \quad \epsilon_F\ll 2\alpha p_F
\end{eqnarray}
where $p_F$ is the momentum at the Fermi surface. In the combination of the disorder broadening and spin-orbit splitting also we can define two different regime depending on which one dominates. The first one is the diffusive regime, corresponding to a high impurity concentration, i.e., $(\alpha/v_f)\epsilon_F\tau\ll1$. The second
regime, which occurs at $(\alpha/v_f)\epsilon_F\tau\gg1$, goes beyond the diffusive
limit and describes the opposite situation of a relatively low
concentration of impurities, where the spin-relaxation time is
close to $\tau_0$. For short-range impurities, when the intraband of the relaxation matrix is neglected, and we can use Eq.~\ref{tmm} to derive the relaxation time in the presence of RSOC. By inserting Eq.~\ref{tmm} in Eq.~\ref{tmm}, and taking the integration over momentum, we receive~\cite{aminsu2,amincubic}
\begin{eqnarray}
	\frac{1}{\tau_{\pm}}\equiv\mathcal{K}_{\pm\bk,\pm\bk'}\approx\frac{1}{\tau_0}\left(\frac{m^{\star}}{1\pm \frac{m^{\star}}{k}}\right)
\end{eqnarray}
In a single matrix in a 2D we can write the relaxation time as 
\begin{eqnarray}
	\hat{\tau}=\tau_0 \left(\sigma^0-\frac{m^\star\alpha}{k}\sigma^z\right)
\end{eqnarray}
By adding above equation inside Eq.~(\ref{p1diagonal}), we have 
\begin{eqnarray}
	\rho_{\bk}^{(-1),\pm \pm}=-\frac{e{\bf E}\cdot\hat{\bk}}{m^\star} \tau_{\pm}k_{\pm}\delta(\omega^{\pm}_{k}-\omega_F),
\end{eqnarray}
where we have used 
\begin{eqnarray}
	\frac{\partial f_0(\omega^{m'}_{\bk'})}{\partial \omega^{m'}_{\bk'}}=-\delta(\omega^{\pm}_{k}-\omega_F)
\end{eqnarray}
and to the lowest order in the spin-orbit splitting, for the momentum we have 
\begin{eqnarray}
	k_{\pm}=k (1\pm \frac{\alpha m^\star}{ k}).
\end{eqnarray}
Hence we have 
\begin{eqnarray}
	\rho_{\bk}^{(-1),mm}=\frac{em^\star\alpha{\bf E}\cdot\hat{\bk}\tau_0}{k}\sigma^z\left(\delta(k-k_F)-k\frac{\partial}{\partial{k}}\delta(k-k_F)\right)\nn \\
\end{eqnarray}
In the eigenstate basic, the spin operator along $y$-direction is given by $s_y=-1/2(cos\theta \sigma^z+sin\theta \sigma^y)$. When the electric field applied along ${\bf E}=E\hat{x}$,  the expectation value of the spin polarization $s^y$ is obtained as 
\begin{eqnarray}
	\langle s^y \rangle= \textnormal{Tr}[s^y \rho_{\bk}^{(-1),mm}]=
	 -N_0e\alpha \tau_0 E_x
\end{eqnarray} 
where  the trace symbol includes the summation over the spin indices and momentum. This is the standard term for the spin polarization reported by Ref.~\cite{edelestein1990}.

\section{CONCLUSIONS}\label{conclusion}
We have derived a general theory of evaluating the second order response of a crystal to an electric field accounting for both interband and intraband contributions. Our theory automatically includes momentum-space Berry phase effect which is often discussed in the context of transport theory. We have derived a quantum kinetic equation by applying the Winger transformation to the quantum Liouvile equation. From this equation, we have obtained that the intraband response is derived by the band diagonal part of the density matrix proportional to the strengthen of disorder potential responsible for properties like longitudinal conductivity. As the system confinement to the Born approximation, the relaxation time matrix is proportional to the disorder potential to leading order. Interband response is captured by off-diagonal part of the density matrix and has sizable contributions to the repose of a crystals. The interplay between diagonal and off-diagonal parts of the density matrix is important in determining the ``guage invariant'' operator Eq.~\ref{p_inter} which is a macroscopic polarization potential to calculate the macroscopic current-density operator $\textbf{J}(t)=d{\textbf P}/dt$. This analysis we presented here can be played essential role in future, more accurate calculations of nonlinear optical response.

\nocite{*}

\bibliography{bpve}
\end{document}